\documentclass[twocolumn,showpacs,preprintnumbers,amsmath,amssymb]{revtex4}

\usepackage{graphicx}
\usepackage{dcolumn}
\usepackage{bm}
\usepackage{epsfig}

\begin{document}

\preprint{Submitted to Phys. Rev. B }

\title{Tunable pinning of a superconducting vortex a by a magnetic vortex }
\author{Gilson Carneiro}
\affiliation{Instituto de F\'{\i}sica, Universidade Federal do Rio de Janeiro,  
C.P. 68528, 21941-972, Rio de Janeiro-RJ, Brasil }
 \email{gmc@if.ufrj.br}

\date{\today}

\begin{abstract}

The interaction  between a straight vortex line in a superconducting film and a soft magnetic 
nanodisk in the magnetic vortex state  in the presence of a magnetic field applied parallel to the film surfaces is studied theoretically. The superconductor is described by  London theory  and the nanodisk by the  
Landau-Lifshitz continuum theory of magnetism, using  the approximation known as  the rigid vortex model. Pinning of the vortex line by the nanodisk is found to result, predominantly,  from the interaction between the vortex line and the changes in the nanodisk magnetization induced  by the magnetic field of the  vortex line and applied field. In the context of the rigid vortex model, these  changes result from the displacement of the magnetic vortex. This displacement is calculated analytically by minimizing the energy, and the pinning potential is obtained. The applied field can tune the pinning potential by  controlling the displacement of the magnetic vortex. The nanodisk magnetization curve is predicted to change in the presence of  the vortex line. 

\end{abstract}
\maketitle


\section{introduction}
\label{sec.int}

Pinning of vortices in superconducting films by arrays of soft nanomagnets placed in the vicinity of the film is a subject that has not been much studied in the literature. Most experimental  \cite{rev1} and theoretical  \cite{coff,wei,sah,myp1,myp2,pri} work carried out so far deals with permanent nanomagnets. For both soft and permanent nanomagnets, pinning results from the action of the inhomogeneous magnetic field created by their magnetization  in the superconductor.  However, in the case of soft nanomagnets, the magnetization depends on the magnetic fields acting on it, including the field caused by the vortex itself. The question is then how this modifies the pinning potential. Recently, this question was investigated for vortices interacting with arrays of soft magnetic nanodisks   using a very simple model in which  the nanodisks are approximated  by  point magnetic dipoles with magnetic moments free to rotate \cite{gmc1}. This model was suggested by  the results of Cowburn et. al. \cite{ckaw}, which show that in small nanodisks the magnetic state is a single domain one, with the magnetization of each nanodisk  behaving  like a giant magnetic moment free to rotate. The calculations of  Ref.\onlinecite{gmc1} show that the pinning potential differs considerably from that for a permanent dipole, and that it can be tuned by a magnetic field applied parallel to the film surfaces. Similar properties are expected for the pinning potential due to soft nonomagnets in general. This paper  studies the pinning of vortices by  nanodisks in another magnetic state: the magnetic vortex. This state is found in larger nanodisks \cite{ckaw,dsk1,dsk2} , and is characterized by the magnetization vector  circulating around a nucleus of small dimensions compared with the nanodisk radius.  The motivation to consider this particular state is that it contains the basic physics of nanomagnets in general, but its theoretical description can be approximated by a simple analytic model. 
Here, the interaction between one straight vortex line and one nanodisk is calculated  using the Landau-Lifschitz  continuum theory of magnetism \cite{ll} for the nanodisk and  London theory for the superconductor \cite{gmc2}. The magnetic vortex state is described by the rigid vortex model introduced by Usov and Peschany \cite{up}, and by Guslienko and collaborators \cite{gus}. These authors obtained the nanodisk equilibrium magnetization  using the variational principle to minimize the energy.  Usov and Peschany \cite{up} proposed an analytic expression for the trial magnetization at zero applied field, with the magnetic vortex core at the nanodisk center, and the core radius as the variational parameter. Guslienko and collaborators \cite{gus} extended  it to finite  fields parallel to the nanodisk faces. They  assumed that the only effect of the applied field is to displace the magnetic vortex rigidly, with the core moving away from the nanodisk center. The new  trial magnetization is the zero field one displaced rigidly with the core, and the core displacement vector is a variational parameter.  The predictions of the rigid vortex model are in reasonable  agreement with numerical calculations  for small displacements compared to the nanodisk radius \cite{up,ckaw,gus}. 

This paper assumes the validity of the rigid vortex model for the nanodisk placed in the vicinity of the superconducting film. The total energy of the nanodisk-superconductor system is obtained exactly as a function of the magnetic vortex displacement vector. The calculations carried out in the paper assume  small displacements compared to the nanodisk radius, and aproximate the the total energy by its expansion to second order in the displacement. The equilibrium displacement of the magnetic vortex is obtained by minimizing the approximate total energy. It is found that, contrary to what happens for the isolated nanodisk, the equilibrium displacement is a non-linear function of the  vortex line and applied fields.  
The pinning potential for the vortex line is also calculated from the approximate total energy. 
The most important contributiont to the pinning potential is found to come from the changes in the  magnetization caused by the displacement of the magnetic vortex. There is also a  contribution from the interaction between the vortex line and the magnetization in the magnetic vortex core, but it is  negligible, due to the small dimensions of the core. It is found that the pinning potential can be tuned  by the applied field. The mechanism is that the field controls the magnetic vortex displacement which, in turn,  modifies the pinning potential. Applications to systems of experimental interest are considered by appropriate choice of the model pararmeters.  The results suggest that the pinning potential for these systems can be estimated by applying the linear response theory of elementary magnetism to the nonodidk. The magnitude of the applied field necessary to annihilate the magnetic vortex is estimated,  from which  changes in the  nanodisk magnetization curve caused by the vortex line can be aticipated.

This paper is organized as follows. The theory for the nanodisk-superconductor system is developed in 
 Sec.\ \ref{sec.pnp}. First, in  Sec.\ \ref{sec.ind}, the rigid vortex model for the isolated nanodisk is briefly reviewed. Then, in  Sec.\ \ref{sec.nsi}, the  nanodisk in the vicinity of the seperconductor is considered,  and the main results of the paper are derived. Finally, in  Sec.\ \ref{sec.dis}, the application of the model to systems of experimental interest is considered, and the conclusions of the paper are stated. The  Appendix gives the mathematical expressions needed in the calculations carried out in Sec.\ \ref{sec.nsi}, and a brief review of their derivation.
%
\begin{figure}[b]
\centerline{\includegraphics[scale=0.3]{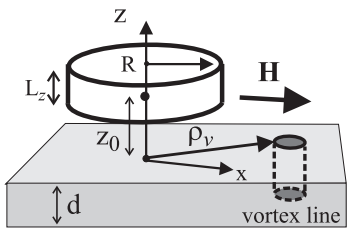}}
\vspace{5mm}
\caption { Schematic view of the superconducting film with one vortex line and a  nanodisk  placed on top. }
\label{fig.fig1}
\end{figure}


\section{pinning potential}
\label{sec.pnp}

The superconductor-nanodisk system is shown schematically in  Fig.\ \ref{fig.fig1}. The superconducting film is planar, of thickness $d$, isotropic and has penetration depth $\lambda$. Its surfaces are parallel to the $x-y$ plane.  One straight vortex line, with vorticity $q=\pm 1$, is present in the film at a position defined by ${\mbox{\boldmath $\rho$}}_{\rm v}$ with respect to an origin in the film top surface. The vortex line core radius is $\xi$.  A thin magnetic nanodisk, of radius  $R$ and thickness  $L_z\ll R$, is located above the film, with the faces parallel to the film surfaces, and  center located above the film at $(x=0,\,y=0,\,z=z_0>0)$. A magnetic field, ${\bf H}$, is applied along the  $x$-axis, that is parallel to the nanodisk faces and to the film surfaces. 
\begin{figure}[t]
\centerline{\includegraphics[scale=0.25]{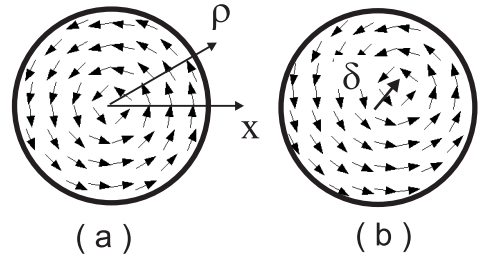}}
\vspace{5mm}
\caption {  Nanodisk magnetization in the magnetic vortex state ($q_m=1$) with core at nanodisk center (a), and  displaced by ${\mbox{\boldmath $\delta$}}$ from nanodisk center (b). Also shown in a) is the cylindrical coordinate system with origin at nanodisk center used in Eq.\ (\ref{eq.rgv}).}
\label{fig.fig2}
\end{figure}

\subsection{isolated nanodisk}
\label{sec.ind}

First, the rigid vortex model for the isolated nanodisk is briefly reviewed. The 
magnetization is written as ${\bf M}({\bf r})=M_s \hat {\bf s}({\bf r})$, $M_s$ being the saturation magnetization. The  nanodisk energy is given by \cite{ll}
\begin{eqnarray}
& & E_D=\frac{M^2_s\,R^2_0}{2}\int\, d^{3}r\sum_{\alpha=x,y,z}
\mid{\mbox{\boldmath $\nabla$}}s_{\alpha}({\bf r})\mid^2 \nonumber \\
& & +\frac{1}{2}\int\, d^{2}r\,\int\, d^{2}r'\, \frac{\sigma({\bf r})\sigma({\bf r}')}{\mid{\bf r} 
-{\bf r}'\mid}
- {\bf m}\cdot{\bf H}\;.
 \label{eq.edk}
\end{eqnarray}
In Eq.\ (\ref{eq.edk}), the first term  is the exchange energy, where $R_0$ is the exchange length. The second is the magnetostatic energy, with  $\sigma({\bf r})=M_s\hat {\bf s}({\bf r})\cdot \hat {\bf n}$ being the density of magnetic charges at the surface of the nanodisk, and $\hat {\bf n}$ is the unit vector in the direction normal to the nanodisk surface. 
The last term is the energy of interaction with  the external field, with ${\bf m}=\int\,d^3r\,{\bf M}({\bf r})$.
It is assumed that the nanodisk is isotropic, and that there are no volume magnetic charges, that is 
${\mbox{\boldmath $\nabla$}}\cdot {\bf M}=0$. The rigid vortex model is a variational minimization of Eq.\ (\ref{eq.edk}).  The trial magnetization is written 
in the  cylindrical coordinate system with origin at the nanodisk center and $z$-axis perpendicular to the nanodisk faces shown in Fig.\ \ref{fig.fig2}a. For $H=0$ Usov and Peschany \cite{up} proposed the following expression 
 \begin{eqnarray} 
  \hat {\bf s}({\mbox{\boldmath $\rho$}}) & = & q_m\, {\hat {\mbox{\boldmath $\theta$}}} \;\; (a<\rho\leq R) \nonumber \\ 
  \hat {\bf s}({\mbox{\boldmath $\rho$}}) & = & q_m\frac{2a\rho}{a^2+\rho^2}{\hat {\mbox{\boldmath $\theta$}}} 
\pm \frac{a^2-\rho^2}{a^2+\rho^2} {\hat {\bf z}} \;\; (0<\rho\leq a) \;,
   \label{eq.rgv}
  \end{eqnarray}
where ${\hat {\mbox{\boldmath $\theta$}}}= {\hat {\bf z}}\times {\hat {\mbox{\boldmath $\rho$}}}$. This equation describes the magnetization vector curling  around the nanodisk center ( Fig.\ \ref{fig.fig2}a ), with  
counterclockwise ( clockwise)  rotation for  $q_m=+ 1\; (-1)$. The variational parameter $a$ is the radius of the magnetic vortex core. The magnetization outside the core is parallel to the nanodisk faces. Inside it has a $z$-component, which can be either positive or negative. The equilibrium value of $a$,  obtained by Usov and Peschany, is $a\approx 0.7(R^2_0L_z)^{1/3}$ \cite{up}.  What is important for the purposes of this paper is not the precise value of $a$, but the fact that it is small compared to $R$. This follows from the assumption that  $R_0,\;L_z \ll R$. 
For ${\bf H}\neq 0$ the trial magnetization proposed  by Guslienko and collaborators \cite{gus} is that  in  Eq.\ (\ref{eq.rgv}) displaced rigidly by ${\mbox{\boldmath $\delta$}}$  from the nanodisk center, that is 
$ {\hat {\bf s}}(\mid {\mbox{\boldmath $\rho$}}-{\mbox{\boldmath $\delta$}}\mid )$ (  Fig.\ \ref{fig.fig2}b ). The equilibrium value of ${\mbox{\boldmath $\delta$}}$ is determined by minimizing the energy. The vortex core radius is unchanged  as long as the displaced core remains within the nanodisk, that is for $\delta < R-a$.  The simplest treatment of the rigid vortex model assumes that the magnetic vortex displacement is small,  and obtains  $E_{D}$ to second order in 
$\delta / R$. The result  is \cite{gus} 
\begin{equation}  
E_{D}({\mbox{\boldmath $\delta$}})= E_{D}(0) +M^2_sV_D[\;\frac{\delta^2_x+\delta^2_y}{2\,\chi\, R^2} -
 q_m\frac{H \delta_y}{M_sR}\;]\;,
  \label{eq.enmd}
  \end{equation} 
where $V_D=\pi R^2L_z$ is the nanodisk volume, and $\delta_x$ and $\delta_y$ are, respectively,  the components of {\mbox{\boldmath $\delta$}} parallel and perpendicular to ${\bf H}$. In Eq.\ (\ref{eq.enmd}) the term quadratic in $\delta$   comes from both the exchange and magnetostatic energies, whereas the term linear in $\delta$ comes from the interaction  with the applied field. The constant $\chi$ is the nanodisk linear magnetic susceptibility ,  and is given by\cite{gus} 
 \begin{equation}  
 \frac{1}{\chi} = -\frac{R^2_0}{R^2} + 2\frac{L_z}{R}[\ln{\frac{8R}{L_z}}-\frac{1}{2}]\;.
  \label{eq.chi}
  \end{equation}
The first term in Eq.\ (\ref{eq.chi}) is the contribution from the exchange energy, and is small because $R\gg R_0$. The second term comes from the magnetostatic energy of the  magnetic  charge density  generated at the nanodisk edges by the vortex displacement. The density of these charges is given by
 \begin{equation}  
 \sigma  =  -q_mM_s\frac{(\hat {\bf z}\times {\mbox{\boldmath $\delta$}})\cdot \hat {\mbox{\boldmath $\rho$}}}
{(R^2-2R {\mbox{\boldmath $\delta$}}\cdot \hat {\mbox{\boldmath $\rho$}}+\delta^2)^{1/2}} 
  \label{eq.sig}
  \end{equation}
The interaction between the magnetic charges in the vortex core and in the nanodisk edges is neglected. To calculate the magnetostatic energy  to second order in $\delta/R$, it is sufficient to use $\sigma$ to first order, that is 
 \begin{equation}  
 \sigma \approx \sigma^{(1)} = -q_mM_s\frac{(\hat {\bf z}\times {\mbox{\boldmath $\delta$}})\cdot \hat {\mbox{\boldmath $\rho$}}}{R} \;.
  \label{eq.sig1}
  \end{equation}
The energy of interaction  with the applied field also comes from $\sigma^{(1)}$. The magnetic moment generated by the displacement of the magnetic vortex is related to $\sigma$, Eq.\ (\ref{eq.sig}), by
\begin{equation}  
 {\bf m}({\mbox{\boldmath $\delta$}})  =  \int \, d^2 r \,\sigma \,{\mbox{\boldmath $\rho$}} \;,
  \label{eq.msg}
  \end{equation}
where the integral is over the nanodisk edge surface. To  first order in $\delta/R$, the magnetic moment is  obtained by replacing $\sigma$ by $\sigma^{(1)}$ in Eq.\ (\ref{eq.msg}). The result is 
is 
 \begin{equation}  
 {\bf m}^{(1)}({\mbox{\boldmath $\delta$}})  =  -q_mM_sV_D\frac{(\hat {\bf z}\times {\mbox{\boldmath $\delta$}})}{R}  \;.
  \label{eq.m1}
  \end{equation}
By symmetry $ {\bf m}({\mbox{\boldmath $\delta$}})$ does contain  terms of second order in  $\delta/R$. 
The equilibrium vortex displacement, obtained by minimizing $E_{D}$ with respect to ${\mbox{\boldmath $\delta$}}$, is 
\begin{equation}
{\mbox{\boldmath $\delta$}}_{eq}  = 
 q_m\chi\frac{\hat {\bf z}\times {\bf H}}{M_s}  \;.
\label{eq.dtid}
\end{equation}
In equilibrium, the magnetic moment is 
\begin{equation}
{\bf m}_{eq}={\bf m}^{(1)}({\mbox{\boldmath $\delta$}}_{eq})=\chi V_D{\bf H}\;.
\label{eq.meq}
\end{equation}
This result shows that  $\chi$ is the nanodisk susceptibility. According to Eq.\ (\ref{eq.dtid}), ${\mbox{\boldmath $\delta$}}_{eq}$ is perpendicular to ${\bf H}$, and depends only on $H$ scaled by $M_s/\chi$. As will be shown in Sec.\ \ref{sec.nsi}, $M_s/\chi$ is also the  scale for the combined effects of the applied field and vortex line field on the magnetic vortex.  
The approximation to second order in $\delta/R$ to $E_D$ is valid only for $\chi H/M_s < 1$. However, it has been used outside this range to estimate the vortex annihilation field,  $H_{an}$, at which the magnetic vortex is destroyed, and the nanodisk  switches to the saturated state \cite{gus}. This estimate assumes that at $H=H_{an}$  the magnetic vortex reaches the nanodisk edge, that is $\delta_{eq}=R$. Thus, according to  Eq.\ (\ref{eq.dtid}), $H_{an}=M_s/\chi$. In this case the nanodisk magnetic moment is the saturation moment, $M_sV_D$ in the direction of ${\bf H}$ (see Eq.\ (\ref{eq.meq})). This estimate for $H_{an}$ is in reasonable agreement with experiments and numerical simulations \cite{gus,ckaw}.

\subsection{nanodisk-superconductor interaction}
\label{sec.nsi}

Now the nanodisk is assumed to be  in the  proximity of  the superconducting film, as shown in  Fig.\ \ref{fig.fig1}. The equilibrium displacement of the magnetic vortex is recalculated by minimizing the total energy of the superconducting film-nanodisk system, $E_{T}({\mbox{\boldmath $\rho$}}_{\rm v},{\mbox{\boldmath $\delta$}})$, to second order in $\delta/R$.  In the London limit \cite{gmc2}
 \begin{equation}  
E_{T}({\mbox{\boldmath $\rho$}}_{\rm v},{\mbox{\boldmath $\delta$}}) =  E_{D}({\mbox{\boldmath $\delta$}})+
E_{MS}({\mbox{\boldmath $\delta$}})+E_{VM}({\mbox{\boldmath $\rho$}}_{\rm v},{\mbox{\boldmath $\delta$}}) \;,
  \label{eq.et1}
  \end{equation} 
where $E_{MS}$ is the energy of interaction of the nanodisk  with the superconducting film in the absence of vortices, and $E_{VM}$  is the interaction energy of the vortex line and the nanodisk.  
The influence of ${\bf H}$ on the superconductor is neglected. This is justifiable  as long as $H$ is smaller than the  lower critical field parallel to the film surfaces. 

\begin{figure}[t]
\centerline{\includegraphics[scale=0.3]{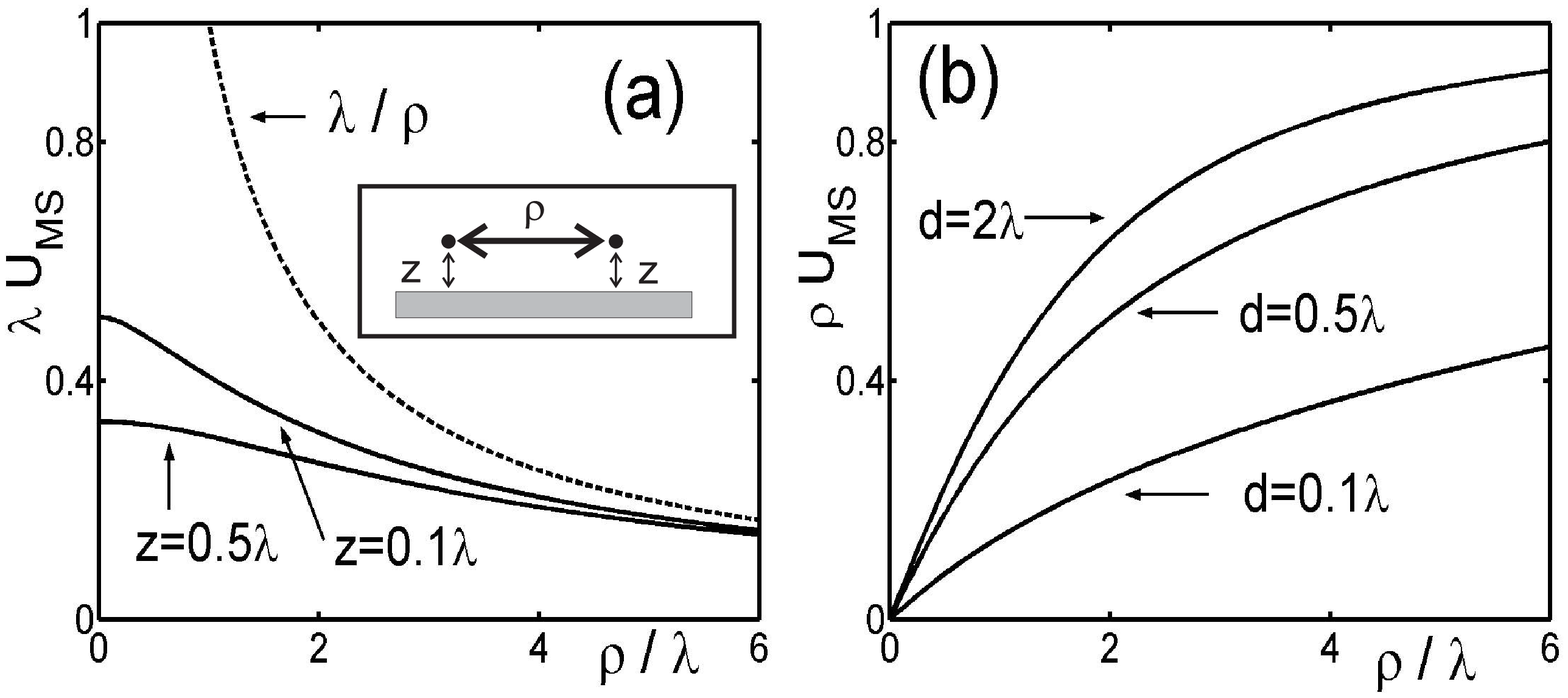}}
\vspace{5mm}
\caption {a) Interaction potential induced by superconducting film between two magnetic charges separated by $\rho$, and at same height $z$,  compared with the Coulomb potential.  a) $\lambda U_{MS}$ vs. $\rho$ for $d=\lambda$ and two $z$ values. Inset: magnetic charges above superconducting film.  b) Ratio $U_{MS}$ to Coulomb potential,  $\rho U_{MS}$, vs. $\rho$ for $z=0.2\lambda$ and several $d$.  }
\label{fig.fig3}
\end{figure}

The energy $E_{MS}$ results from the interaction between the nanodisk and the magnetic field of the screening current  generated by  it in the superconductor\cite{gmc2}. As shown in the Appendix, $E_{MS}$ can be written as an interaction between magnetic charges, 
\begin{equation}
E_{MS} = \frac{1}{2}\int d^{2}r\int d^{2}r'\, \sigma({\bf r})\sigma({\bf r}')
U_{MS}({\bf r};{\bf r}')\; .
 \label{eq.ems} 
\end{equation}
Thus, in the presence of the superconducting film, the interaction between  magnetic charges is modified   by the addition of $U_{MS}$ to the Coulomb potential ( see Eq.\ (\ref{eq.edk})). In  Fig.\ \ref{fig.fig3}, $\;U_{MS}$ is shown  for two magnetic charges separated by $\rho$ and at the same height $z$, and compared with the Coulomb potential.   For large distances, $U_{MS}$ coincides with the Coulomb potential.  Large distances meaning $\rho \gg \Lambda= 2\lambda^2/d\;$  for thin films ($d\ll \lambda$) and $\rho \gg \lambda$ for films with  $d\sim \lambda$. For short distances, $U_{MS}$ is considerably smaller than the Coulomb potential. The effects of $E_{MS}$ on the magnetic vortex are to change  the vortex core radius, $a$, and nanodisk susceptibility, $\chi$. However these  modifications are small. The reason is that the dominant contributions to $a$ and $\chi$ come from short distances \cite{up,gus}, where $U_{MS}$  is much smaller than the Coulomb interaction.  For instance, the contribution to $\chi^{-1}$  from  $E_{MS}$ is found to be about one order of magnitude smaller than that from the Coulomb interaction, Eq.\ (\ref{eq.chi}). Hereafter $E_{MS}$ is neglected.

The energy $E_{VM}$ is  given by \cite{gmc2} 
  \begin{eqnarray}  
E_{VM}&=& - \int\, d^3r' \, {\bf M}({\bf r'})\cdot\, {\bf b}({\bf r}'-{\mbox{\boldmath $\rho$}}_{\rm v}) 
 \label{eq.evma}
\\
& = &\int\, d^2r' \, \sigma({\bf r'}) \Phi({\bf r}'-{\mbox{\boldmath $\rho$}_{\rm v}}) \,.
 \label{eq.evmb}
  \end{eqnarray}
where ${\bf b}({\bf r}) = -\,{\mbox{\boldmath $\nabla$}}\,\Phi({\bf r})$ is the magnetic field created by the vortex line outside the film. 
There are   two contributions to $E_{VM}$: one from the magnetic vortex core, and another from the magnetic charge density at the nanodisk edges, $\sigma$, Eq.\ (\ref{eq.sig}), denoted by $E^{(e)}_{VM}$.  The contribution to $E_{VM}$ from the  core is negligible, as justified later.   In order to obtain $E^{(e)}_{VM}$ to  second order in $\delta/R$ it is necessary to use $\sigma$ to the same order. The first order term is $\sigma^{(1)}$, Eq.\ (\ref{eq.sig1}). The second order term is, according to Eq.\ (\ref{eq.sig}), given by  
  \begin{equation}  
 \sigma^{(2)}= \sigma^{(1)}\frac{{\mbox{\boldmath $\delta$}}\cdot \hat {\mbox{\boldmath $\rho$}}}{R}  \;\;.
 \label{eq.sig2}
  \end{equation}

The contribution from $\sigma^{(1)}$ to $E_{VM}$, denoted $E^{(e1)}_{VM}$, is obtained by noting that $\sigma^{(1)}$ can be interpreted as resulting from the uniform magnetization 
 \begin{equation} 
{\bf M}^{(1)}=-q_mM_s\frac{\hat {\bf z}\times {\mbox{\boldmath $\delta$}}}{R} \;.  
 \label{eq.bm1} 
 \end{equation}
Using  Eq.\ (\ref{eq.evma}) it follows that  
 \begin{equation}  
E^{(e1)}_{VM}= - V_D{\bf M}^{(1)}\cdot\, {\bf B}_{\perp}(-{\mbox{\boldmath $\rho$}}_{\rm v}) \,,
 \label{eq.upm}
  \end{equation}
where 
  \begin{eqnarray}  
 {\bf B}_{\perp}(-{\mbox{\boldmath $\rho$}}_{\rm v}) & = & \frac{1}{V_D}\,\int_{\rm disk}\, d^3r' \, {\bf b}_{\perp}({\bf r}'-{\mbox{\boldmath $\rho$}}_{\rm v}) \nonumber \\
& = & -qB_{\perp}(\rho_{\rm v})\hat {\mbox{\boldmath $\rho$}}_{\rm v},
 \label{eq.bav}
  \end{eqnarray}
is the average over the nanodisk volume of the component of the vortex field perpendicular to the $z$-direction ( parallel to the nanodisk faces ), ${\bf b}_{\perp}$. The argument of $ {\bf B}_{\perp}$ in  Eq.\ (\ref{eq.bav}) is $-{\mbox{\boldmath $\rho$}}_{\rm v}$   because  ${\mbox{\boldmath $\rho$}}_{\rm v}$ is defined with respect to the origin shown in Fig.\ \ref{fig.fig1}, whereas in the definition of the vortex line field  the origin is at the vortex line ( see Eq.\ (\ref{eq.phi})). Thus 
\begin{equation}  
E^{(e1)}_{VM} =q_m\,M_sV_D\,qB_{\perp}(\rho_{\rm v})\,\frac{\delta_T}{R}
\;\;,
 \label{eq.evme1}
  \end{equation}
where $\delta_T$ denotes the component of ${\mbox{\boldmath $\delta$}}$ perpendicular to ${\bf B}_{\perp}(-{\mbox{\boldmath $\rho$}}_{\rm v})$ ( parallel to $-{\hat {\bf z}}\times {\hat {\mbox{\boldmath $\rho$}}}_{\rm v}$, see Fig.\ \ref{fig.fig4}).
The contribution from  $\sigma^{(2)}$ is calculated in the Appendix. The result is  
  \begin{equation}  
E^{(e2)}_{VM}=  q_m\,M_sV_D  \, qB_1(\rho_{\rm v})\,\frac{\delta_L\delta_T}{R^2}   \;\; ,
 \label{eq.evme2}
  \end{equation} 
where $\delta_L$ denotes the component of ${\mbox{\boldmath $\delta$}}$ parallel to 
${\bf B}_{\perp}(-{\mbox{\boldmath $\rho$}}_{\rm v})$ (Fig.\ \ref{fig.fig4}), and $B_1(\rho_{\rm v})$ has  dimension of  magnetic field, and is given by Eq.\ (\ref{eq.b1}). 
\begin{figure}[t]
\centerline{\includegraphics[scale=0.2]{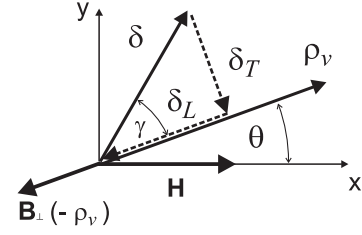}}
\vspace{5mm}
\caption {Definition of the displacements $\delta_L$,  $\delta_T$, and  angles $\theta$, and $\gamma$. }
\label{fig.fig4}
\end{figure}

Now the equilibrium displacement of the magnetic vortex is calculated neglecting the contribution from the vortex core to 
$E_{VM}$.  The total energy  is thus 
\begin{eqnarray}  
& &E_{T}({\mbox{\boldmath $\rho$}}_{\rm v},{\mbox{\boldmath $\delta$}}) = E_{D}(0) + 
M^2_sV_D[\;\frac{\delta^2_T+\delta^2_L}{2\,\chi \,R^2}+ \nonumber \\ 
& & q_m\frac{qB_{\perp}(\rho_{\rm v})\delta_T-H\delta_y}{M_s R} + 
 q_m\, \frac{qB_1(\rho_{\rm v})\delta_L\delta_T}{M_sR^2}\;] \;\;.
 \label{eq.etmv} 
  \end{eqnarray}
The components of   ${\mbox{\boldmath $\delta$}}$ parallel and perpendicular to ${\bf H}$, $\delta_x$ and $\delta_y$, respectively, are related to $\delta_L$ and $\delta_T$  by  (see Fig.\ \ref{fig.fig4}).
 \begin{eqnarray}
\delta_x & = &-\delta_L \cos{\theta} +\delta_T \sin{\theta} \nonumber \\
\delta_y &= & -\delta_L \sin{\theta} -\delta_T \cos{\theta} \; .
\end{eqnarray}
Minimizing $E_T$ with respect to ${\mbox{\boldmath $\delta$}}$, with the vortex line held fixed at ${\mbox{\boldmath $\rho$}_{\rm v}}$,   it follows that 
 \begin{eqnarray} 
& & \frac{\delta_{T,eq}({\mbox{\boldmath $\rho$}_{\rm v}})}{R}   =   q_m\frac{q{\tilde B}_{\perp}-{\tilde H}\cos{\theta}
+ q_mq{\tilde B}_1{\tilde H}\sin{\theta}}{1-{\tilde B}_1^2} \; ,\nonumber \\
& &\frac{\delta_{L,eq}({\mbox{\boldmath $\rho$}_{\rm v}})}{R}  =  \ -q_m\frac{{\tilde H}\sin{\theta}+q_mq{\tilde B}_1
(q{\tilde B}_{\perp}-{\tilde H}\cos{\theta})}{1-{\tilde B}_1^2} \; ,\nonumber \\
& & {\tilde B}_{\perp} \equiv  \frac{\chi B_{\perp}(\rho_{\rm v})}{M_s}\; , {\tilde B}_{1}  \equiv  \frac{\chi B_{1}(\rho_{\rm v})}{M_s}\; , {\tilde H} \equiv  \frac{\chi H}{M_s}\; .
 \label{eq.dlt1}
  \end{eqnarray}  
Thus ${\mbox{\boldmath $\delta$}}_{eq}$ depends only on the vortex line  and  applied fields scaled by $M_s/\chi$, and is non-linear in these fields. The non-linearity results from the inhomogeneity of the vortex field over the nanodisk volume, which is responsible for $B_1$ being non-zero. This solution is only valid if ${\tilde B}_{1}^2<1$. Otherwise ${\mbox{\boldmath $\delta$}}_{eq}$, Eq.\ (\ref{eq.dlt1}), does not correspond to a minimum of  $E_{T}({\mbox{\boldmath $\rho$}}_{\rm v},{\mbox{\boldmath $\delta$}})$, Eq.\ (\ref{eq.etmv}).  The consequences of Eq.\ (\ref{eq.dlt1}) are discussed in detail in Sec.\ \ref{sec.dis}.

Now the neglect of the core contribution to $E_{VM}$, denoted $E^{(c)}_{VM}$, is justified. Since $a$ is small compared to $R$,  $E^{(c)}_{VM}$ can be estimated by approximating the core by a point dipole located at the nanodisk center with the total magnetic moment of the core.  According to Eq.\ (\ref{eq.rgv}), it  has only the $z$-component  
$m^{(c)}_z=\pm M_s\pi a^2L_z(2\ln{2}-1)$. Thus, $E^{(c)}_{VM}\approx  - m^{(c)}_z\,b_z({\mbox{\boldmath $\delta$}}-{\mbox{\boldmath $\rho$}_{\rm v}}+z_0\hat{\bf z})$.  
The basic reason why $E^{(c)}_{VM}$ can be neglected is that  it is proportional to $a^2$, which is already a small quantity. One effect of  $E^{(c)}_{VM}$ is to modify the vortex core. Since it depends on $a$, its contribution must be added to $E_D$ in order to obtain the equilibrium value of $a$. However, the effect is a small one because 
$E_D\sim M^2_sa^4/L_z$ \cite{up}, so that  $E^{(c)}_{VM}/E_{D}\sim (L_z/a)^2(b_z/M_s)$, and $(b_z/M_s)$ is small, as shown in Sec.\ \ref{sec.dis}. Another effect of $E^{(c)}_{VM}$ is to displace the magnetic vortex from the nanodisk center. The displacement can be estimated as follows. The force exerted by the superconducting vortex on the magnetic vortex core is $\mid {\mbox{\boldmath $\nabla$}}E^{(c)}_{VM}\mid \sim M^2_sV_D (a/R)^2 b_z/M_s\ell$, where $\ell$ is the typical scale for variations of $b_z$, namely $\ell \sim z_0$ for thin films ($d \ll \lambda$) and $\ell \sim \lambda$ for films with $d\sim \lambda$. This force must be balanced by the elastic force $M_s^2V_D \delta/\chi\,R^2$, which gives $\delta/R\sim (a/R)^2 (R/\ell) \chi \,b_z/M_s$, and is smaller than the vortex displacement caused by $E^{(e)}_{VM}$ at least by a factor $a/R$.

The pinning potential for the vortex line is defined as the equilibrium total energy for the vortex line held fixed at 
${\mbox{\boldmath $\rho$}}_{\rm v}$. That is 
$ U_p({\mbox{\boldmath $\rho$}}_{\rm v})= E_{T}({\mbox{\boldmath $\rho$}}_{\rm v},{\mbox{\boldmath $\delta$}}_{eq})$.
Thus 
 \begin{eqnarray}  
& & U_p({\mbox{\boldmath $\rho$}}_{\rm v})= - \frac{\chi V_D}{2(1-{\tilde B}_1^2)}[B^2_{\perp}- 2qB_{\perp}H\cos{\theta} \nonumber \\
& & +2q_mq{\tilde B}_1H\sin{\theta}(qB_{\perp}-H\cos{\theta})]\; .
  \label{eq.umvh}
  \end{eqnarray}  
This result shows that the pinning potential can be tuned by the  applied field. The mechanism is that, according to 
Eq.\ (\ref{eq.dlt1}), ${\bf H}$ controls the equilibrium displacement of the magnetic vortex which, in turn, modifies 
$U_p$.

In order to evaluate the consequences of the above results  it is necessary to attribute values to the model parameters.
This is done next.

\section{discussion}
\label{sec.dis}

The  results of Sec.\ \ref{sec.nsi} are applicable to arrays of soft magnetic nanodisks  placed on top of 
low-T$_c$ superconducting films, provided that the nanodisks are sufficiently far apart to neglect  dipole-dipole interactions between them, and the vortex density is low enough to neglect vortex-vortex interactions. Realistic values for the model parameters are the following \cite{rev1,ckaw,dsk1,dsk2}. Permalloy nanodisks: $100\,nm < R < 500\,nm$,  
$L_z \sim 10\, nm$,  $M_s = 0.8\,kG$, and $R_0\, \sim 15\, nm$. Superconducting films at zero temperature: 
$\lambda = 75\, nm$, $\xi = 0.1\, \lambda $, $d \sim\, 0.2 - 2\, \lambda$. With this value for $\lambda$,  the  scale for the vortex fields  is $\phi_0/\lambda^2= 3.6\, kG$.  The distance from the nanodisk center to the film surface is chosen as the smallest possible, namely $z_0=0.2\lambda=2\xi$. This   value takes into account the existence of an insulating layer between the film and the nanodisk, of thickness $\sim \xi$,  to avoid the proximity effect. 
It is found that for these parameter values the vortex fields $B_{\perp}$ and $B_1$ are small compared to $M_s/\chi$. As  consequences, ${\tilde B}_1$ is negligible in Eq.\ (\ref{eq.dlt1}), the magnetic  vortex displacement is  small compared with and nanodisk radius, and its dependence on  the vortex field is linear. This is discussed in detail next. 
\begin{figure}[t]
\centerline{\includegraphics[scale=0.32]{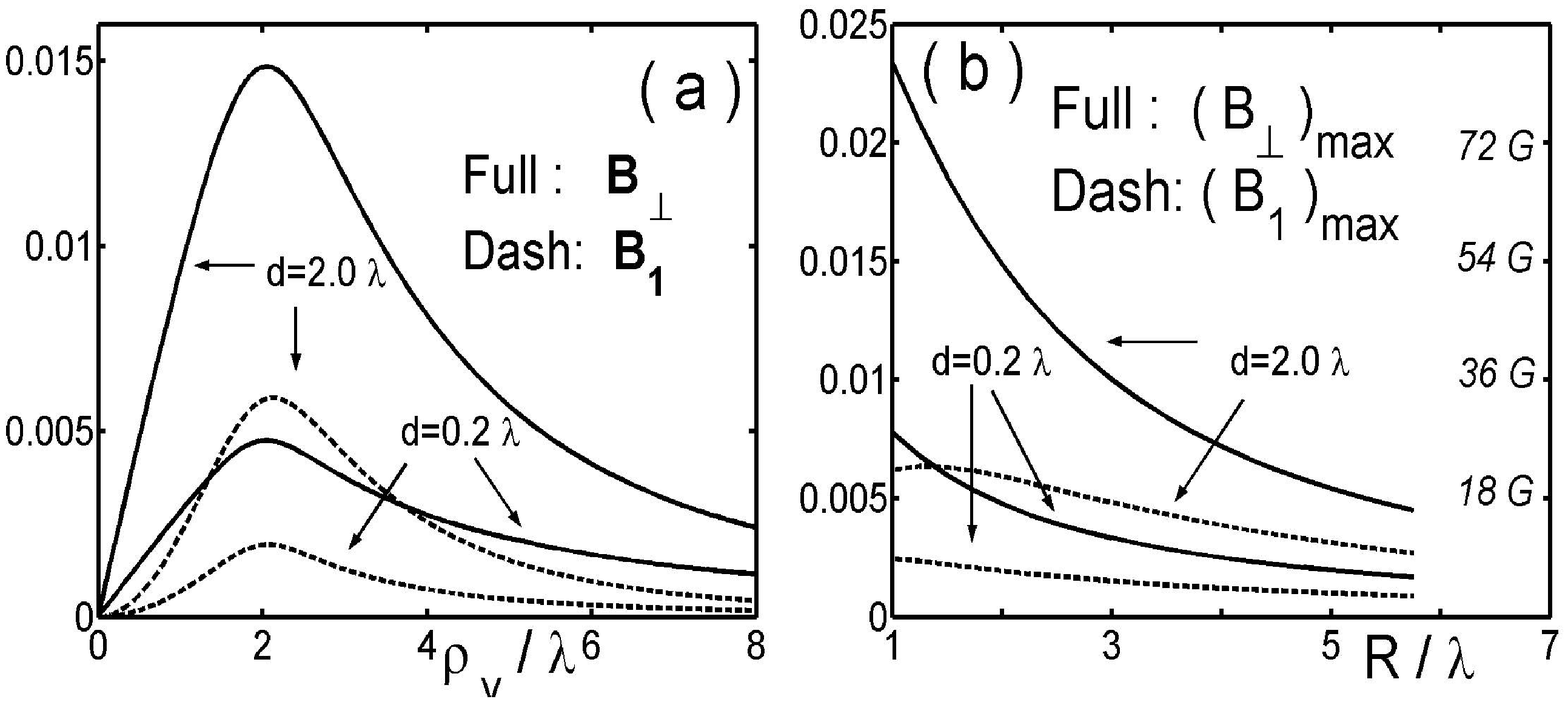}}
\vspace{5mm}
\caption{ a) Fields $B_{\perp}$  and $B_1$ , in units of $\phi_0/\lambda^2$, vs. $\rho_{\rm v}$ for $R=2.0\lambda$.  b) Maximum values of $B_{\perp}$  and $B_1$ in units of $\phi_0/\lambda^2$ vs. $R$. The right-hand scale in Gauss corresponds to left-hand one for $\lambda= 75\,nm$. Parameters:  $ \xi=0.1\lambda, \,z_0=0.2\lambda,\, L_z=0.1\lambda$.}
\label{fig.fig5}
\end{figure}

The fields $B_{\perp}$ and $B_1$ are shown in Fig .\ \ref{fig.fig5}. They  depend only on scaled variables, with $\lambda$ as the length scale and $\phi_0/\lambda^2$ as the magnetic field scale (see Eqs.\ (\ref{eq.phi}), (\ref{eq.f1k})).
Their dependencies  on  $\rho_{\rm v}$ are shown in Fig.\ \ref{fig.fig5}.a. Both vanish for $\rho_{\rm v}=0$, and have a maximum at $\rho_{\rm v}\sim R$, with $B_1$ smaller than $B_{\perp}$ by a factor $\sim 3$. The maximum values of  $B_{\perp}$ and $B_1$ depend on $R$ as shown as   in Fig.\ \ref{fig.fig5}.b. The right-hand scale in Gauss in  this figure corresponds to the left-hand side one for  $\phi_0/\lambda^2=3.6 \, kG$.
Both $B_{\perp}$ and $B_1$ increase non-linearly with $d$, up to $d \sim 2.0 \lambda$. For larger $d$, they change little, because  the  vortex line field is generated by currents flowing close to the film surface.
These results  indicate that ${\tilde B}_1$ is  small. Using   $\chi \sim 2$, which corresponds to $R\sim 200\, nm,\, L_z=10\, nm$ ( see Eq.\ (\ref{eq.chi}) ), the maximum values of ${\tilde B}_1$, obtained from the $B_1$ data in Fig .\ \ref{fig.fig5}b,  are $({\tilde B}_{1})_{max} \sim 5 \times 10^{-2}$ for $d=2.0\lambda$ and $({\tilde B}_{1})_{max} \sim 1.0 \times 10^{-2}$ for  $d=0.2\lambda$.  Thus ${\tilde B}_1^2 \ll 1$, and can be neglected in the denominators of Eqs.\ (\ref{eq.dlt1}) and  (\ref{eq.umvh}). 
The terms with ${\tilde B}_{1}$ in numerator of these equations are also small, because they involve the combinations ${\tilde B}_{1}{\tilde H}$, and ${\tilde B}_{1}{\tilde B}_{\perp}$, which are in general smaller than the other terms. 
\begin{figure}[t]
\centerline{\includegraphics[scale=0.3]{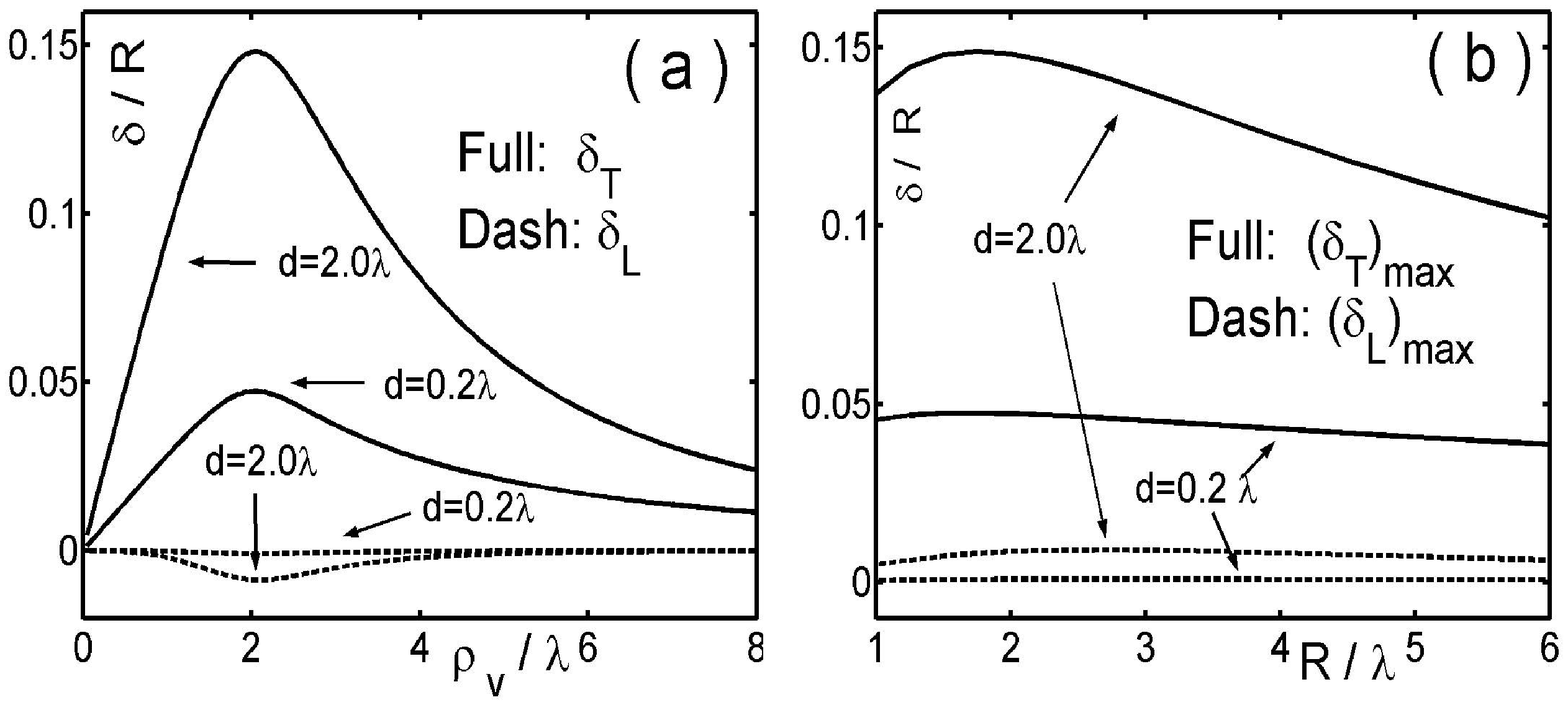}}
\vspace{5mm}
\caption{  a) Equilibrium displacements of the magnetic vortex ($q_m=1$) for , $H=0$, $R=2.0\lambda$. b) Maximum values of the equilibrium displacements vs. $R$ for $H=0$. Parameters:  $ \xi=0.1\lambda, \,z_0=0.2\lambda,\, L_z=0.1\lambda$,  $M_s=0.22\phi_0/\lambda^2$.  }
\label{fig.fig6}
\end{figure}

Results for $\delta_{L,eq}$ and $\delta_{T,eq}$ at $H=0$ are shown in Fig.\ \ref{fig.fig6}. In this case,  according to  Eq.\ (\ref{eq.dlt1}), $\mid \delta_{L,eq}/\delta_{T,eq}\mid =  {\tilde B}_{1} $. Thus, $\mid \delta_{L,eq}/\delta_{T,eq}\mid \ll 1$ since $ {\tilde B}_{1}$ is small. 
The curves in Fig.\ \ref{fig.fig6}.a show  this. These plots also show that $\delta_{T,eq}/R <0.15 $  for $d=2.0\lambda$ and  $\delta_{T,eq}/R <0.06 $ for $d=0.2\lambda$. In Fig.\ \ref{fig.fig6}.b the maximum values of  $\delta_{L,eq}$ and $\delta_{T,eq}$ are shown as a function of $R$. Both vary little with $R$. The reason is that, according to Eq.\ (\ref{eq.dlt1}), the $R$-dependence of $\delta_{L,eq}$ and $\delta_{T,eq}$ is only through the products  $\chi B_{\perp}$ and $\chi B_1$. These are nearly independent of $R$, because there is a cancelation between the increase of $\chi$ with $R$ ( see Eq.\ (\ref{eq.chi}))  and the decrease of  $B_{\perp}$ and $B_1$  with $R$ ( Fig.\ \ref{fig.fig5}.b). 
 
\begin{figure}[h]
\centerline{\includegraphics[scale=0.3]{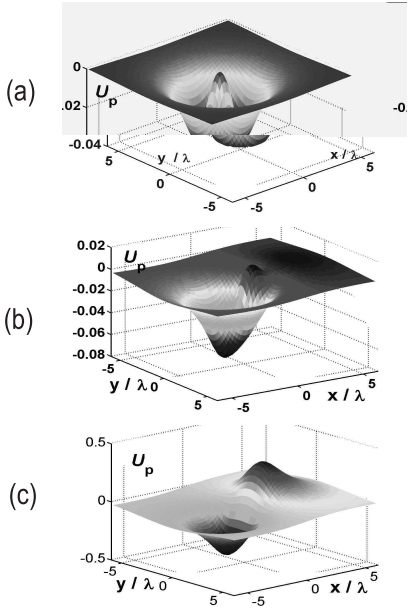}}
\vspace{5mm}
\caption { Pinning potential for a vortex line by a magnetic nanodisk in the vortex state, in units of $\epsilon_0 \lambda=\phi_0^2/(16\pi^2\lambda)$, for $ d=2.0\lambda,\,z_0=0.2\lambda,\,\xi=0.1\lambda,\, R=1.5\lambda,\, L_z=0.1\lambda, \,M_s=0.1\phi_0/\lambda^2$.
a) $H=0$. b) $H=0.01\phi_0/\lambda^2$. c) $H=0.1\phi_0/\lambda^2$.}
\label{fig.fig7}
\end{figure}

The above discussion suggest  that, for the parameter values described above,   a good  approximation to the equilibrium displacement is to  put ${\tilde B}_{1}$  equal to zero in Eq.\ (\ref{eq.dlt1}). In this case  ${\mbox{\boldmath $\delta$}}_{eq}$, is identical to that for the isolated nanodisk in the applied field  ${\bf H}_T ={\bf B}_{\perp}(-{\mbox{\boldmath $\rho$}}_{\rm v}) + {\bf H}$. That is,  Eq.\ (\ref{eq.dlt1}) with ${\tilde B}_{1}=0$, is identical to  Eq.\ (\ref{eq.dtid}) with ${\bf H}$ replaced by ${\bf H}_T$.
In this approximation the pinning potential, Eq.\ (\ref{eq.umvh}),  reduces to
 \begin{equation}  
U_p= - \frac{\chi V_D}{2}[B^2_{\perp}- 2qB_{\perp}H\cos{\theta}] \;.
  \label{eq.uml}
  \end{equation}  
This result is, up to a constant,  the magnetostatic energy of interaction between the magnetic moment induced by ${\bf H}_T$ in the nanodisk, ${\bf m}_{eq}= \chi V_D{\bf H}_T $,  and ${\bf H}_T$  itself. That is 
 \begin{equation}  
U_p= - \frac{1}{2}{\bf m}_{eq}\cdot {\bf H}_T  + \frac{\chi V_D H^2}{2}  \;.
  \label{eq.umlb}
  \end{equation}
This simple result is the just the linear response approximation from elementary macroscopic magnetism \cite{ecm} applied to  the nanodisk. It depends only on the nanodisk susceptibility, $\chi$,  and on the macroscopic vortex field acting on it, ${\bf B}_{\perp}(-{\mbox{\boldmath $\rho$}}_{\rm v})$. 
Results for the pinning potential $U_p({\mbox{\boldmath $\rho$}}_{\rm v})$, based on  Eq.\ (\ref{eq.uml}), are shown in  Fig.\ \ref{fig.fig7} as two-dimensional plots for  characteristic values of $H$. 
For $H=0$, $U_p$ has circular symmetry, and is the same for vortices ($q=1$) and anti-vortices ($q=-1$), with a degenerate minimum located in a circle of radius  $\rho_{\rm v}\sim R$ (Fig.\ \ref{fig.fig7}a). For $H\neq 0$,  $U_p$ has a non-degenerate minimum and also a maximum along the direction of ${\bf H}$. The plot in Fig.\ \ref{fig.fig7}b corresponds to $H\lesssim B_{\perp}$,  whereas that in Fig.\ \ref{fig.fig7}c is for $H > B_{\perp}$. In the latter case the pinning potential reduces to $U_p({\mbox{\boldmath $\rho$}}_{\rm v})= qV_D\chi B_{\perp}(\rho_{\rm v}) H \cos{\theta}$, which is identical to that for a nanodisk with permanent uniform  magnetization $\chi {\bf H}$. Since $B_{\perp} \ll M_s/\chi$, the range of $H$ values for which the vortex line field influences the pinning potential is small. Thus, except for small $H$, the spatial dependence of the pinning potential is given by $B_{\perp}(\rho_{\rm v}) \cos{\theta}$, and  the magnitude is proportional to $H$.  This allows for continuous tuning of $U_p$ over a wide range. 
It is useful to compare the magnitude of $U_p$ to that for the pinning potential for a nanodisk with permanent magnetization, equal to the saturation one, $U\sim M_sV_DB_{\perp}$. For $H<M_s/\chi$, $U_p$ is smaller than $U$, since  for $H\lesssim B_{\perp}$,  $\mid U_p/U\mid \sim \chi B_{\perp}/M_s \ll 1$ , and for $M_s/\chi> H > B_{\perp}$, $\mid U_p/U\mid \sim \chi H/M_s < 1$. Using Eq.\ (\ref{eq.umlb}) to extrapolate $U_p$ to the region  $H>M_s/\chi$, it follows that $U_p\sim U$ only at  $H\sim M_s/\chi$.  

The linear response approximation is now used  to estimate the effect of the vortex line  on the magnetic vortex annihilation field, $H_{an}$. As in Sec.\ \ref{sec.ind}, $H_{an}$ is estimated as the value of $H$ for which $\delta_{eq}=R$. In the linear response approximation this occurs for  $H_T = M_s/\chi$. Since $B_{\perp} \ll M_s/\chi$, $H_{an}$ differs little from that predicted for isolated magnetic vortex, that is   $H_{an}\sim M_s/\chi$. Consequently, the nanodisk magnetization curve is expected to differ little from that for the isolated nanodisk.
%

In summary then, for the parameter values described earlier, the main conclusions are: i) the vortex line field causes only a small  displacement of the magnetic vortex, ii) the pinning potential is the energy of interaction  between  the magnetic moment induced in the  nanodisk by the vortex line and applied fields and the fields themselves, calculated  according to the laws of macroscopic linear magnetism, iii) the vortex line has little effect on the nanodisk magnetization curve.    

\begin{figure}[b]
\centerline{\includegraphics[scale=0.32]{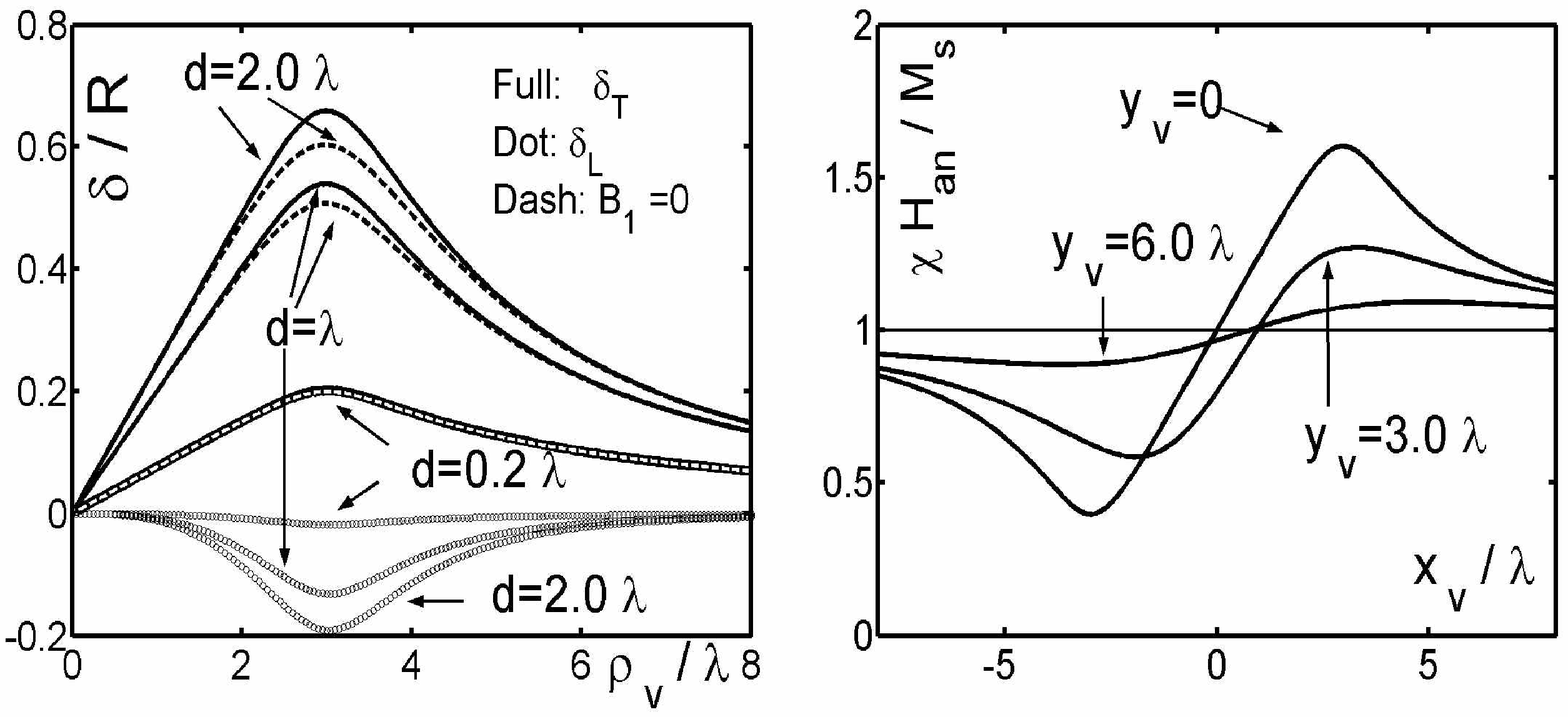}}
\vspace{5mm}
\caption { a) Equilibrium displacements of the magnetic vortex ( $q_m=1$ ) for $H=0$. b) Magnetic vortex annihilation field for vortex line at position $(x_{\rm v}, y_{\rm v})$ vs. $x_{\rm v}$ for $y_{\rm v}$ constant. Parameters:  $\xi=0.1\lambda, \,z_0=0.2\lambda,\,R=3.0\lambda\, L_z=0.1\lambda, \, M_s=0.05\phi_0/\lambda^2$.}
\label{fig.fig8}
\end{figure}

Now the non-linear dependence of ${\mbox{\boldmath $\delta$}}_{eq}$ on the vortex line field and  applied fields, predicted by Eq.\ (\ref{eq.dlt1}), is considered. The question is whether or not there are parameter values for which 
${\tilde B}_{1}$ is sufficiently large that the non-linearity is important and what are its consequences. One limitation on  ${\tilde B}_{1}$ is that  Eq.\ (\ref{eq.dlt1}) only makes sense if $\delta/R<1$. This effectively restricts ${\tilde B}_{1}$ to relatively small values. One reason is that the denominators in  Eq.\ (\ref{eq.dlt1})  cannot be too small. Another reason is that $B_1$ and $B_{\perp}$ are  not independent  (see Eqs.\ (\ref{eq.bavb}), (\ref{eq.b1})), so that if ${\tilde B}_{1}$ is not sufficiently  small, ${\tilde B}_{\perp}$  is large enough to make 
$\delta/R>1$. It is found that, in general, $B_1<B_{\perp}$, as  exemplified  in   Fig.\ \ref{fig.fig5}. To investigate this quantitatively,  the equilibrium displacement of the magnetic vortex for  $H=0$ is calculated using a new set of model parameters.
These are chosen in order to give displacements larger than the ones obtained with the parameters mentioned at the beginning of this Section.  For the superconducting film the parameters are the same as before ($\xi=0.1 \lambda, \; z_0=0.2\lambda=2\xi$), except for $\lambda$ which is now chosen as $\lambda=50\,nm$. In this case the fields $B_{\perp}$ and $B_1$ are unchanged in units of $\phi_0/\lambda^2$, but their values in Gauss change because now $\phi_0/\lambda^2=8\,kG$. The new parameters for the nanodisk  are chosen as $R=3.0 \lambda,\; L_z= 0.1 \lambda, M_s=0.4 kG$. In this case $\chi=3.0$ and $M_s/\chi =0.13\, kG$. 
The results for the magnetic vortex displacements  are shown in Fig.\ \ref{fig.fig8}a, and compared with those obtained from Eq.\ (\ref{eq.dlt1}) with ${\tilde B}_{1}=0$. The differences between the curves for $\delta_{T,eq}$ with ${\tilde B}_{1}\neq 0$ and ${\tilde B}_{1}=0$ are due to ${\tilde B}_{1}^2$ in the  denominator of Eq.\ (\ref{eq.dlt1}). The results show that the differences is small, even when $\delta_{T,eq}$ is a considerable fraction of $R$. For $\delta_{L,eq}$, only the curves for  ${\tilde B}_{1}\neq 0$ are shown, because $\delta_{L,eq}$ vanishes for  ${\tilde B}_{1}=0$.  The non-zero  $\delta_{L,eq}$ shown in Fig.\ \ref{fig.fig8}b results from the non-linearity of  Eq.\ (\ref{eq.dlt1}), with the   most important contribution coming from the numerator.  Comparing  the curves for $\delta_{L,eq}$ and $\delta_{T,eq}$ in  Fig.\ \ref{fig.fig8}a,  and using  $\mid \delta_{L,eq}/\delta_{T,eq}\mid = {\tilde B}_{1} $, it follows that ${\tilde B}_{1}\lesssim 0.3$. This shows that the validity of Eq.\ (\ref{eq.dlt1}) requires small values of  ${\tilde B}_{1}$. 

The non-linear effects in the pinning potential $U_p$,  Eq.\ (\ref{eq.umvh}), for $H=0$ come only from   
${\tilde B}_{1}^2$ in the denominator, and are also small according to the above discussion. The properties of $U_p$ for the new set of parameters are similar to those described earlier.

The  results shown in  Fig.\ \ref{fig.fig8}.a also have  consequences for the magnetic vortex annihilation field, 
$H_{an}$. When the  vortex line   displaces  the magnetic vortex by a significant fraction of the nanodisk radius it takes only a small applied field to annihilate the magnetic vortex by further displacing it to the nanodisk edge. In this case   $H_{an}$  differs significantly from that for the isolated nanodisk.  To estimate $H_{an}$, Eq.\ (\ref{eq.dlt1}) with  ${\tilde B}_{1}=0$  is used. In this case, as discussed above, the magnetic vortex is annihilated, when $H_T\sim M_s/\chi$. This estimate for $H_{an}$ is shown in Fig.\ \ref{fig.fig8}.b. A strong dependence of $H_{an}$ on the vortex line position results. The absolute minimum of $H_{an}$ occurs when  the vortex line is at the pinning potential minimum for $H=0$, located at $y_{\rm v}=0$ and $x_{\rm v}\sim -R$, since in this case the displacement of the magnetic vortex is maximum. Thus, when the vortex line is in equilibrium with the nanodisk $H_{an}$ is minimum. This result indicates that the vortex line in equilibrium with the nanodisk can significantly  modify the  magnetization curve.

To conclude then, the magnitude of the displacement of the magnetic vortex caused by the vortex line depends on the particular values of the model parameters. In the cases were the displacement is a significant fraction of the nanodisk radius, effects of the non-linear relationship between the displacement and the vortex line field are felt, but are small, and the magnetization curve for the nanodisk in equilibrium with the vortex line is predicted to differ from that for the isolated nanodisk.  In all cases, the  displacement of the magnetic vortex and the pinning potential can be estimated by applying the linear response theory of elementary magnetism to the nanodisk. This suggests that the pinning potential for nanomagnets with other geometrical forms for which the magnetic vortex state has been reported \cite{vts} can  be likewise estimated.   

\acknowledgments

Research supported in part by the Brazilian agencies CNPq, CAPES, FAPERJ,  and FUJB.

\appendix*

\section{London theory results}
\label{sec.mtd}

Here some results of London theory for the superconducting film, obtained  in Refs.\onlinecite{gmc2,gcehb}, are reviewed. The objective is to write out the mathematical expressions needed to carry out the calculations mentioned in Sec.\ \ref{sec.nsi}.

\subsection{nanodisk-screening current interaction}
\label{sec.ns}

In the arrangement shown in Fig.\ \ref{fig.fig1}, the nanodisk generates in the  superconducting film  a screening current which, in turn, creates a magnetic field at the nanodisk. The energy of interaction of the nanodisk  with this field is given by
\begin{equation}  
E_{MS}= -\frac{1}{2} \int\, d^3r \, {\bf M}({\bf r})\cdot\, {\bf b}_{sc}({\bf r}) \,,
 \label{eq.emsb}
  \end{equation}
where ${\bf b}_{sc}({\bf r})$ is the field of the screening current outside the film.
The field ${\bf b}_{sc}$ is can be written as 
\begin{eqnarray}
{\bf b}_{sc}({\bf r}) &= & -{\mbox{\boldmath $\nabla$}}\Phi_{sc}({\bf r})\; , \nonumber \\
\Phi_{sc}({\bf r}) & = &
\int d^3r' ({\bf M}({\bf r}')\cdot\, {\mbox{\boldmath $\nabla$}'})
 U_{MS}({\bf r};{\bf r}') \,,
 \label{eq.psc}
  \end{eqnarray}
where $U_{MS}$ is given by
\begin{eqnarray}
& & U_{MS}({\bf r};{\bf r}')=\int\, d^2k\, 
e^{i{\bf k }\cdot ({\bf r}_{\perp}-{\bf r}'_{\perp})}\,e^{-k(z+z')}\;g(k)\; , \nonumber \\
& & g(k) =  \frac{\sinh{\tau d}}{k[e^{-\tau d}(k-\tau)^2 -e^{\tau d}(k+\tau)^2]} \; ,\nonumber \\
\tau&=&\sqrt{k^2+\lambda^{-2}}\;,
 \label{eq.ums}
  \end{eqnarray}
where ${\bf r}_{\perp}$ denotes the component of ${\bf r}$ perpendicular to the $z$-direction. Assuming that ${\mbox{\boldmath $\nabla$}}\cdot {\bf M}=0$, and integrating by parts, Eq.\ (\ref{eq.emsb}) can be written as Eq.\ (\ref{eq.ems}).

\subsection{vortex line field}
\label{sec.vtf}

The field created by the  vortex line outside the film is given by 
\begin{eqnarray}
 {\bf b}({\bf r})& = &-\,{\mbox{\boldmath $\nabla$}}\,\Phi({\bf r})\;\;,  \nonumber \\
 \Phi({\bf r})& = &-q\frac{\phi_0}{\lambda^2}\int\,\frac{d^{2}k}{(2\pi)^2}\,e^{i{\bf k }\cdot{\bf r}_{\perp}}\;e^{-kz}\, F_1(k)\;\;, 
 \label{eq.phi}
\end{eqnarray}
where ${\bf r}$ is the position vector with respect to an origin at the vortex line, and 
\begin{eqnarray}
 F_1(k) =  
\frac{e^{-2\xi^2k^2}[(k+\tau)e^{\tau d}+(k-\tau)e^{-\tau d}-2k]}
{k\tau[(k-\tau)^2e^{-\tau d}-(k+\tau)^2e^{\tau d}]}\;. 
 \label{eq.f1k}
\end{eqnarray}
Integrating Eq.\ (\ref{eq.phi}) over the direction of 
${\bf k}$,  $\Phi$ can be written as 
\begin{equation}
\Phi( r_{\perp},z)=-\,q\int^{\infty}_0\,\frac{dk}{2\pi}\,k J_0(kr_{\perp})\,e^{-kz}\, F_1(k)\; ,
 \label{eq.phb}
\end{equation}
where $J_0$ is the Bessel function of first kind. Thus, the vortex field can be written as
\begin{eqnarray}
{\bf b}({\bf r})&=& b_{\perp}(r_{\perp},z)\,\hat {\bf r}_{\perp}+b_z(r_{\perp},z)\,\hat{\bf z}\;,\nonumber\\
b_{\perp}&=&-q\frac{\phi_0}{\lambda^2}
\,\int^{\infty}_0\,\frac{dk}{2\pi}\,k^2 J_1(kr_{\perp})\,e^{-kz}\, F_1(k)\;\; , \nonumber\\
b_z&=&-q\frac{\phi_0}{\lambda^2}
\,\int^{\infty}_0\,\frac{dk}{2\pi}\,k^2 J_0(kr_{\perp})\,e^{-kz}\, F_1(k)\;. 
 \label{eq.bvl} 
\end{eqnarray}
Using  Eq.\ (\ref{eq.bvl}) and  Eq.\ (\ref{eq.bav}), it follows that 
\begin{eqnarray}
B_{\perp}(\rho_{\rm v})&=&-\frac{\phi_0}{\lambda^2}
\,\int^{\infty}_0\,\frac{dk}{2\pi}\,S_1(k)\,k^2 J_1(k\rho_{\rm v})\,e^{-kz_0}\, F_1(k)\;\; , \nonumber\\
S_1(k)&=&\frac{2J_1(kR)}{kR}\,\frac{2\sinh{(kL_z/2)}}{kL_z}\;. 
\label{eq.bavb}
\end{eqnarray}

\subsection{superconducting vortex-magnetic vortex interaction} 
\label{sec.svmv}

Here the contribution from $\sigma^{(2)}$ to the vortex line - nanodisk interaction  is calculated. Using Eqs.\ (\ref{eq.evmb}), and (\ref{eq.phi}), $E^{(e2)}_{VM}$ can be written as 
  \begin{eqnarray}   
& & E^{(e2)}_{VM} =  \int^{z_0+L_z/2}_{z_0-L_z/2}\, dz' \, \int^{2\pi}_0 d\theta'\, \sigma^{(2)}(\theta') 
\nonumber \\
& & (-q\frac{\phi_0}{\lambda^2})\int\,\frac{d^{2}k}{(2\pi)^2}\,
e^{i({\mbox{\boldmath \small$\rho$}'}-{\mbox{\boldmath \small$\rho$}_{\rm v})}\cdot{\bf k }}\;e^{-kz}\, F_1(k)\;,
 \label{eq.evme2b}
  \end{eqnarray} 
where $\mid {\mbox{\boldmath \small$\rho$}'}\mid =R$.   
According to  Eq.\ (\ref{eq.sig2}), $\sigma^{(2)}$ is given by
\begin{equation}
\sigma^{(2)}(\theta')=q_mM_s(\frac{\delta}{R})^2\,\sin{(\beta'-\theta')}\cos{(\beta'-\theta')}\,
 \label{eq.sig2b} 
\end{equation}
where  $\beta'$ and $\theta'$ are, respectively, the angles between ${\mbox{\boldmath $\delta$}}$ and ${\bf k}$, and between ${\mbox{\boldmath $\rho$}}'$ and ${\bf k}$. 
 Integrating over $z'$, $\theta'$ and $\beta'$, it follows that  
  \begin{equation}  
E^{(e2)}_{VM}=  q_m\,M_sV_D  \, qB_1(\rho_{\rm v})\,\frac{\delta^2\cos{\gamma}\sin{\gamma}}{R^2}   \;\; ,
 \label{eq.evme2c}
  \end{equation} 
where 
  \begin{eqnarray}   
 B_1(\rho_{\rm v}) & = & -\frac{\phi_0}{\lambda^2}\int^{\infty}_0\,\frac{dk}{2\pi}S_2(k) k^2 J_2(k\rho_{\rm v})
 e^{-kz_0}F_1(k) \nonumber \;,\\
 S_2(k)&=&\frac{2J_2(kR)}{kR}\,\frac{2\sinh{(kL_z/2)}}{kL_z}\;,
 \label{eq.b1}
  \end{eqnarray} 
and $\gamma$ is the angle between  ${\mbox{\boldmath $\delta$}}$
 and ${\mbox{\boldmath $\rho$}}_{\rm v}$ (Fig.\ \ref{fig.fig4}). Using $\delta_L=\delta\cos{\gamma}$ and $\delta_T=\delta\sin{\gamma}$, Eq.\ (\ref{eq.evme2c}) is identical to Eq.\ (\ref{eq.evme2}).

\end{document}